# Motion-related Artefact Classification Using Patch-based Ensemble and Transfer Learning in Cardiac MRI

Ruizhe Li[1], Xin Chen[1]

[1]Intelligent Modelling & Analysis Group, School of Computer Science,
University of Nottingham, UK

**Abstract.** Cardiac Magnetic Resonance Imaging (MRI) plays an important role in the analysis of cardiac function. However, the acquisition is often accompanied by motion artefacts because of the difficulty of breath-hold, especially for acute symptoms patients. Therefore, it is essential to assess the quality of cardiac MRI for further analysis. Time-consuming manual-based classification is not conducive to the construction of an end-to-end computer aided diagnostic system. To overcome this problem, an automatic cardiac MRI quality estimation framework using ensemble and transfer learning is proposed in this work. Multiple pre-trained models were initialised and fine-tuned on 2-dimensional image patches sampled from the training data. In the model inference process, decisions from these models are aggregated to make a final prediction. The framework has been evaluated on CMRxMotion grand challenge (MICCAI 2022) dataset which is small, multi-class, and imbalanced. It achieved a classification accuracy of 78.8% and 70.0% on the training set (5-fold cross-validation) and a validation set, respectively. The final trained model was also evaluated on an independent test set by the CMRxMotion organisers, which achieved the classification accuracy of 72.5% and Cohen's Kappa of 0.6309 (ranked top 1 in this grand challenge). Our code is available on Github: https://github.com/ruizhe-l/CMRxMotion.

**Keywords:** Cardiac MRI, Motion Artefacts, Ensemble Learning, Patch-based Classification.

## 1    Introduction

Cardiac Magnetic Resonance Imaging (MRI) is widely used in clinical practice for cardiac function analysis and disease diagnosis. However, during cardiac MRI acquisition, respiratory motions introduce image artefacts, resulting in the inaccurate image-based analysis[1]. Therefore, it is essential to estimate the image quality of cardiac MRI. Manual annotation is time-consuming and laborious for human experts. Therefore, an automatic approach is needed to classify the image quality of cardiac MRI.

Convolutional Neural Network (CNN)-based methods (e.g. ResNet [2] and EfficienNet [3]) have shown excellent performance on natural image quality assessment tasks. Wu et al. extracted high-level features using a CNN-based model to learn a hierarchical feature degradation [4]. Varga et al. employed ResNet that achieved blind image quality assessment without any quality information for pristine reference images [5]. More recently, Vision Transformer (ViT) [6] also achieved remarkable performance on image quality assessment tasks [7].

Compared to natural images, the medical image is more challenging due to the lack of large datasets and the difficulty of annotation. Oksuz et al. proposed a k-space data augmentation strategy and a curriculum learning scheme to detect motion artefacts that achieved a good performance on a large cardiac MRI dataset [8]. Ettehadi et al. proposed a 3D residual Squeeze-and-Excitation CNN method for Diffusion MRI (dMRI) artefact detection [9]. Instead of binary-level classification, they achieved a remarkable result on multiclass classification for multiple artefacts. Besides, various other deep-learning-based approaches achieved acceptable performance on image quality detection and classification tasks in medical images [10][11][12].

However, the performance of the k-space based approaches highly depends on extensive effort and fine-grained design for parameter fitting, which is time costly and infeasible in practice [13]. In addition, most of the methods also rely on a large amount of balanced training data.

The CMRxMotion grand challenge (MICCAI 2022) dataset [14] consists of small, multi-class, and imbalanced data, which poses a huge challenge in training deep-learning-based methods. To overcome these difficulties, we propose a patch-based ensemble learning method based on pre-trained models. Our design principle is threefold, as described below. 1) To address the issue of small data with varying image sizes, each of the 3D volume is split into 2D slices and randomly sampled using image patches with a fixed size. Three different classifiers were trained for ensemble decision making, each of the classifier is fine-tuned on a pre-trained 2D classification model. This patch-based ensemble learning helps to avoid model overfitting. 2) To reduce the impact of large image intensity variations, the image gradient magnitude of the original image is used to combine with the original image for ensemble learning. 3) The majority voting is used to summarise the results from different models for decision making. We evaluated our method using 5-fold cross-validation. It was also tested on an independent validation set provided by the challenge organiser.

## 2    Methods

An overview of the proposed method is shown in **Fig.1**. In the training phase, automatic segmentation of the heart region is firstly performed using an encoder-decoder deep CNN model (i.e. U-Net [15]). Each 2D slice in the input cardiac MRI is considered independently. A gradient magnitude map for each slice is calculated to reduce the appearance difference across images. After that, random sampling is applied to every 2D slice of both the original image and the gradient magnitude map to generate a set of image patches with a fixed size. These image patches are only around the heart region assisted by a heart segmentation model. The ensemble learning technique is then used to train multiple models with different architectures, each model is trained separately on the image patches of the original image and the gradient magnitude map.

In the test phase, image patches are randomly sampled from the original input image and its corresponding gradient magnitude map around the segmented heart region. For every sub-classifier, majority voting is used to make a slice-level decision based on the aggregation of predictive probability values from multiple image patches. Then a biased voting (described in section 3.3) is applied to summarize the slice-level results to get the subject-level results. Finally, another majority voting is applied to produce the final subject-level decisions based on the outputs of all sub-classifiers.

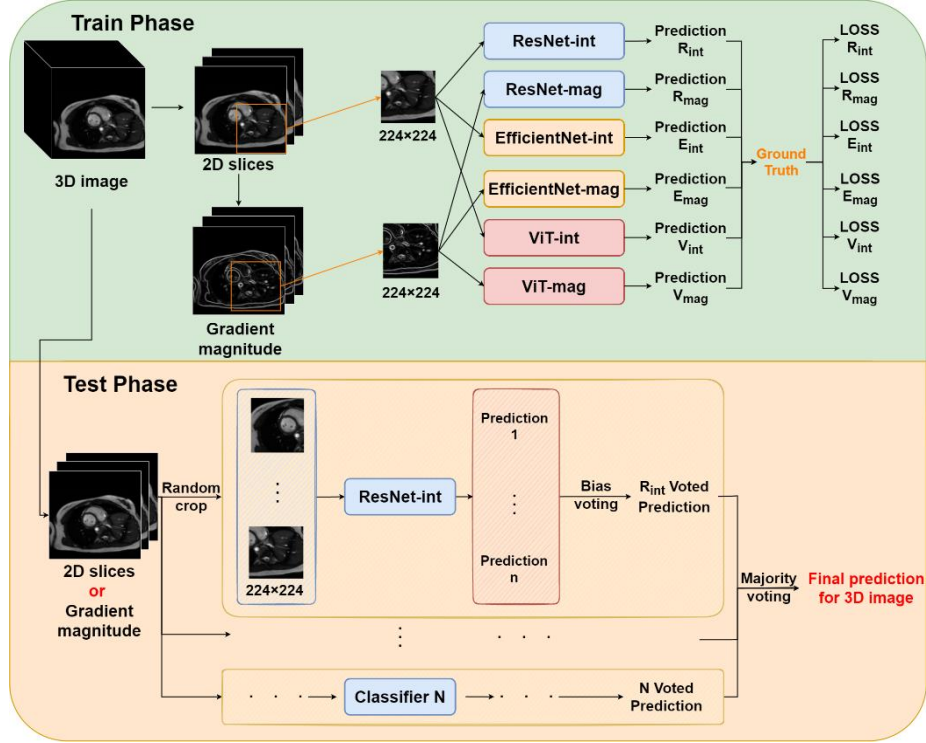

**Fig. 1.** Overview of the proposed framework. In the training phase, three networks (ResNet, EfficientNet and ViT) are trained on 2D image patches of original image and gradient magnitude map respectively, resulting in six classifiers. In the test phase, the input multi-slice image is sampled to multiple random patches. The final decision is voted by all classifiers on all patches.

### 2.1 Image Pre-processing using Gradient Magnitude

The Cardiac MRI data acquired under different postures and breathing patterns (i.e. breath-hold, breath freely, etc.) have intensity and shape discrepancies, which pose challenges for both classification and segmentation tasks. In the context of motion artefact detection, machine learning models could be confused by motion artefacts and intensity changes. Image gradient magnitude is a commonly used method for feature representation[16][17] to reduce intensity variations. The calculation of gradient magnitude on a 2D image is expressed in Equation (1).

$$\nabla f(x,y) = \sqrt{(\frac{\partial f(x,y)}{\partial x})^2 + (\frac{\partial f(x,y)}{\partial y})^2} \qquad (1)$$

where $\nabla f(x, y)$ indicates the gradient magnitude at pixel location $(x, y)$. $\frac{\partial f(x,y)}{\partial x}$ and $\frac{\partial f(x,y)}{\partial x}$ are the gradients in the x and y directions respectively. Prewitt operators (Equation (2)) are used to compute an approximal gradient of the image intensity [18].

$$h_x = \begin{bmatrix} +1 & 0 & -1 \\ +1 & 0 & -1 \\ +1 & 0 & -1 \end{bmatrix} \text{ and } h_y = \begin{bmatrix} +1 & +1 & +1 \\ 0 & 0 & 0 \\ -1 & -1 & -1 \end{bmatrix} \quad (2)$$

The approximal gradient map $\widetilde{\nabla f}$ is calculated by applying 2-dimensional convolutions between the Prewitt operators and the original image in both directions (described in Equation (3)).

$$\widetilde{\nabla f} = \sqrt{(h_x * I)^2 + (h_y * I)^2} \quad (3)$$

where $I$ is the original image and the symbol $*$ denotes the 2-dimensional convolution operation.

## 2.2 Image Patch Sampling

The CMRxMotion dataset used in this paper (described in Section 3.1) is small and diverse. To solve both problems, a patch-based CNN classification model is applied to increase the number of training data and ensure the training data input to the model is in a fixed size [19][20]. Before inputting the data into a classification model, each multi-slice cardiac MRI scan is split into 2D slices and then randomly sampled into small patches, where each patch is classified separately.

We also assumed that the heart region is more important than other image regions for motion artefact detection in cardiac MRI. Hence, we propose to sample the image patches only around the heart region. To achieve this, an automatic cardiac tissue segmentation network is trained using an encoder-decoder network (U-Net [15]). The dataset from CMRxMotion task 2 (described in section 3.1), is used to train this segmentation model. Then, according to the segmentation results, we randomly sample the image patches around the foreground region by ensuring that each image patch contains at least 80% of the foreground region in each image slice. In addition, the patch size is set to 224×224, which matches the input image size of pre-trained models (section 2.3) and ensures good coverage of the heart region.

## 2.3 Transfer-learning-based Model Ensemble

Different deep learning models have their own advantages. An ensemble learning of combining different models may help to improve the accuracy and reliability of the predicted result. Additionally, the use of pre-trained models can help to avoid model overfitting to a small training dataset. In our method, three different networks are selected to construct an ensemble learning framework, including ResNet [2], EfficientNet [3] and ViT [6]. They have different filter sizes, model architectures and learning strategies (i.e. convolution vs. attention) that could potentially diversify the decision making process from different aspects. Rather than training from sketch, the official pre-trained models trained on ImageNet [21] were used to initialise the trainable weights. For each model, to match the input image dimension (1 for our dataset and 3 for the pre-trained model), we add a learnable convolutional layer with 3×3 kernel size at the beginning of each model, followed by a batch normalisation layer and a ReLU activation function.

During the training process, the pre-trained weights of all other layers except for the batch normalisation, the added convolutional layer and the decision-making layers (fully connected layers) are fixed. The training phase is to retrain the normalisation layers to refine it based on the mean and variance on the new domain specific dataset, the added convolutional layer for image feature conversion, and the fully connected layers to refine the decision making.

As illustrated in Fig. 1, the three pre-trained models are trained separately based on image patches of the original image and the gradient magnitude map, resulting in a total of six models in the ensemble framework. In the model inference stage, slice-level decisions are produced based on aggregating (i.e. average) the predictive probability values of each class using all image patches in each slice. A subject-level decision is made by a biased voting of slice-level predictions. Then the final decision is based on majority voting based on the six decisions from the sub-classifiers. In the case of an equal vote, the class that corresponds to a more severe motion artefact is selected.

## 3 Experiments and Results

### 3.1 Dataset

The CMRxMotion challenge Task 1 dataset is used to evaluate the proposed method. It consists of 320 Cardiac MRI scans from 40 volunteers. All the volunteers were asked to act in 4 situations, respectively: a) full breath-hold during acquisition; b) halve the breath-hold period; c) breathe freely; and d) breathe intensively. In each case, images at end-diastole (ED) and end-systole (ES) phases were acquired, resulting in 8 volumes per volunteer. The cardiac MRI scans were in a variety of widths and heights ranging from 400 to 600 pixels, as well as varying numbers of slices from 9 to 13.

The data was divided into 3 classes depending on the level of motion artefacts, 70, 69 and 21 scans for mild, intermediate and severe, respectively. To avoid the validation set and training set being too similar, the data were randomly divided into 5-fold on subject-level. Specifically, each volunteer's scans of 8 phases were within in the same fold for achieving fair cross-validation. In addition, due to the differences of respiratory motions for different volunteers, the data was not able to be split exactly evenly. By multiple random splits, each fold had almost balanced number of scans in mild and intermediate class and at least 3 scans in severe class. Each fold contained 32 scans.

The CMRxMotion challenge Task 2 dataset was used to train the segmentation model. The segmentation labels (left ventricle (LV), right ventricle (RV) and left ventricular myocardium (MYO)) are provided by the organiser. To simplify the segmentation task, all the 3 labels were combined into a single label (foreground) as a binary segmentation problem. We then trained a vanilla U-Net [15] on this simplified dataset.

It achieved an acceptable result with dice coefficient of 0.899. This model was then applied in both the training and testing phases to detect an approximate heart region for image patch sampling.

### 3.2 Batch Balanced Training

In CMRxMotion dataset, the number of images in different classes is very unevenly distributed. Therefore, a batch-based data balancing strategy was designed to balance the data in each training batch. In each batch, the same number of 2D slices were acquired from each class to form a class-balanced batch for model training. Although the images from classes with fewer data appeared multiple times in the same epoch of training, the random patch-based approach was able to significantly reduce the repetition of data by sampling different image regions.

### 3.3 Bias Voting in Testing

**Algorithm 1:** Biased Voting for one subject

```
 1: Input: number of slices predicted as mild, intermediate and severe (N_1, N_2, N_3)
 2:        bias ratio r_1, r_2
 3: IF N_2 + N_3 > r1 × (N_1 + N_2 + N_3) then
 4:    IF N_3 > r2 × (N_2 + N_3)
 5:       return severe
 6:    else
 7:       return intermediate
 8:    end
 9: else
10:    return mild
11: end
```

By analysing the data, we discovered that for each 3D volume, a small number of 2D slices with intermediate or severe motion artefacts, leads to an annotation of intermediate or severe for the subject. This shows that majority voting is not suitable at the slice-level. To solve this problem, we use a slice-based bias voting strategy. As shown in Algorithm 1, for each 3D volume, after the patch-based slice-level prediction, the number of slices predicted as mild, intermediate and severe are counted as $N_1$, $N_2$ and $N_3$ respectively. Then, we set $N_2$ and $N_3$ as a group, when the proportion of this group ($N_2 + N_3$) in total number ($N_1 + N_2 + N_3$) is greater than a ratio $r_1$, the subject-level precision will be chosen between intermediate and severe. Similarly, inside the group, when $N_3$ is greater than a ratio $r_2$, the final prediction will be set as severe. Through cross-validation on the training set, $r_1$ and $r_2$ are set to 0.4 and 0.25 respectively.

### 3.4 Other Parameter Setting

For the training of all classifiers, a fixed learning rate of 0.0001 was used. All models were trained with 10 epochs with Adam as optimiser. Dropout was added for every fully connected layer with a ratio of 0.5 to avoid overfitting. For both training and testing, the batch size was set to 30, where each batch included 10 patches in each class. In the testing phase, the number of randomly sampled patches was 20 for majority voting. The results were predicted from the best checkpoint based on a 5-fold cross-validation.

### 3.5 Experiments and Results

**Table 1.** The results for 6 different classifiers and the ensembled model on the training set over 5-fold cross-validation. Includes overall accuracy and accuracy for each level of motion artefacts (Mild, Intermediate (Inter) and Severe). Cohen's Kappa (Kappa) is also reported to measure the agreement between the ground truth and the prediction [22].

| Classifier | Accuracy (%) | | | | Kappa |
|---|---|---|---|---|---|
| | Overall | Mild | Inter | Severe | |
| ResNet-int | 71.9 | **87.1** | 55.1 | 76.2 | 0.5492 |
| ResNet-mag | 70.6 | 82.9 | 52.2 | **90.5** | 0.5333 |
| EfficientNet-int | 73.8 | 84.3 | 63.8 | 71.4 | 0.5689 |
| EfficientNet-mag | 76.9 | 81.4 | 71.0 | 81.0 | 0.6227 |
| ViT-int | 73.1 | 85.7 | 58.0 | 81.0 | 0.5705 |
| ViT-mag | 73.8 | 71.4 | **75.4** | 76.2 | 0.5720 |
| Ensembled | **78.8** | 85.7 | 69.6 | 85.7 | **0.6564** |

Three different networks (section 2.3) combined with two types of input images (original image and gradient magnitude map) were trained and evaluated separately before model ensemble, which result in six different classifiers named as ResNet-int, ResNet-mag, EfficientNet-int, EfficientNet-mag, ViT-int, ViT-mag. A 5-fold cross-validation was performed on each classifier. A total of 30 classifiers (6 models × 5 fold) were trained on CMRxMotion dataset. The results are reported in **Table 1**. According to **Table 1**, all models achieved 70% to 80% overall accuracies. Their Cohen's Kappa values were all larger than 0.53. The results from EfficientNet trained on gradient magnitude map achieved the highest Kappa value of 0.6227, which was a substantial agreement between the ground truth and the predicted results. However, the performance of different models in different classes varies significantly. ResNet-int and ResNet-mag achieved the best prediction results in mild and severe classes respectively. ViT-mag achieved the best result in intermediate class. This demonstrates the inherent instability of these sub-classifiers due to the insufficiency and imbalance of data.

Therefore, the above findings confirmed the necessity of using an ensembled framework. We combined the results of all models by majority voting (section 2.3). As shown in **Table 1**. The Ensemble method yielded the best Cohen's Kappa, which achieved 0.6564 as a substantial agreement between the ground truth and the prediction. It also achieved the best overall accuracy (78.8%) and remarkable results (85.7%, 69.6% and 85.7%) for each class. The ensemble method not only achieved higher prediction accuracy, but also offered better stability and robustness compared to a single classifier.

The ensemble method achieved a competitive result on the validation dataset provided by the organiser through an online validation system, which achieved a classification accuracy of 70% and Cohen's Kappa of 0.5588. Subsequently, we submitted our code and trained model to the CMRxMotion organisers. Our final model was then independently evaluated on a test set by the CMRxMotion organisers, which achieved the classification accuracy of 72.5% and Cohen's Kappa of 0.6309 (**ranked top 1 in this grand challenge**).

## 4  Conclusions and Discussion

In this paper, we proposed a patch-based classification method based on ensemble and transfer learning for predicting three levels of motion artefacts in cardiac MRI. It was evaluated on CMRxMotion (MICCAI 2022) grant challenge dataset that consists of small, multi-class, and imbalanced data. Through the assistance of a heart segmentation model trained on the segmentation labels from task 2, the image patches were sampled around the heart region where the motion artefacts are more likely to be detected. For model training, the ensemble technique is used to train three state-of-the-art classification networks (ResNet, EfficientNet and ViT) on both the original intensity images and the gradient magnitude maps, resulting in six classification models. The final prediction is formed by majority voting on the outputs of all classifiers. The evaluation results of 5-fold cross-validation on the training set show the ensembled method outperforms all single models. The final model won first place in this grand challenge.

Furthermore, we also try to identify the shortcomings of our method. We noticed that the quality of the slices from the apical and basal regions of cardiac MRI may not contribute significantly to the quality score rated by humans as shown in Fig.2. The subject in Fig. 2 was rated as mild by humans, even the basal and apical slices contain more severe motion artefacts. For the methods like our framework, which is trained on 2D slices, this noisy labelling will interfere with the model's performance. Although patch-based training and ensemble learning can mitigate this effect, the contradictory labels may still confuse the model, leading to inaccurate prediction in certain cases. Therefore, in future work, we may design a method to remove these noisy labels and retrain a model with more representative examples.

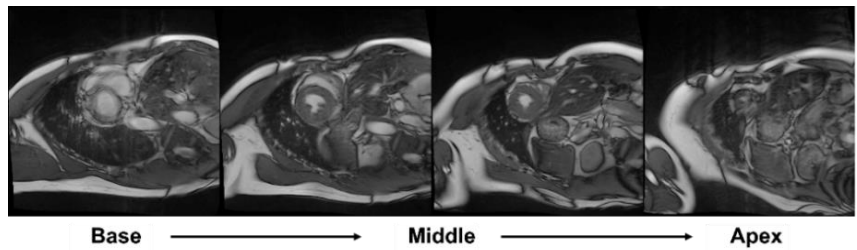

**Fig. 2.** Examples of cardiac MRI slices which motion artefacts level is marked as mild at subject level based on the two middle slices. However, the Basal and Apex slices have more severe motion artefacts.